\documentclass[journal]{IEEEtran}
\usepackage{cite}
   \usepackage[dvips]{graphicx}

\begin{document}
%
\title{Temperature-dependence of detection efficiency in NbN and TaN SNSPD}

\author{Andreas Engel,
  Kevin Inderbitzin,
  Andreas Schilling,
  Robert Lusche,
  Alexei Semenov,
  Heinz-Wilhelm H\"{u}bers,
  Dagmar Henrich,
  Matthias Hofherr,
  Konstantin Il'in,
  and Michael Siegel
\thanks{Manuscript received October 9, 2012.}
\thanks{A. Engel (corresponding author e-mail: andreas.engel@physics.uzh.ch), K. Inderbitzin and A. Schilling are with the Physics Institute, University of Z\"{u}rich, 8057 Z\"{u}rich, Switzerland.}%
\thanks{R. Lusche, A. Semenov and H.-W. H\"{u}bers are with the Institute of Planetary Research, DLR e.V. (German Aerospace Center), 12489 Berlin, Germany.}%
\thanks{D. Henrich, M. Hofherr, K. Il'in and M. Siegel are with the Institute for Micro- und Nanoelectronic Systems, Karlsruher Institute for Technology (KIT), 76187 Karlsruhe, Germany. A.~E., K.~I., and A.~S.\ received support from the Swiss National Science Foundation grant No.\ 200021\_135504/1, and D.~H., M.~H., K.~I., and M.~S.\ received supported in part by DFG Center for Functional Nanostructures under sub-project A4.3.}}%
\markboth{}{\thepage}

\newcommand{\lco}{\ensuremath{\lambda_c}}

\maketitle

\begin{abstract}
  We present systematic measurements of the temperature-dependence of detection efficiencies in TaN and NbN superconducting nanowire single-photon detectors. We have observed a clear increase of the cut-off wavelength with decreasing temperature that we can qualitatively describe with a temperature-dependent diffusion coefficient of the quasi-particles created after photon absorption. Furthermore, the detection efficiency at wavelengths shorter than the cut-off wavelength as well as at longer wavelengths exhibit distinct temperature dependencies. The underlying causes and possible consequences for microscopic detection models are discussed.
\end{abstract}

\IEEEpeerreviewmaketitle

\section{Introduction}
\IEEEPARstart{A}{t the beginning} of the research on superconducting nanowire single-photon detectors (SNSPD) \cite{Goltsman01} the detectors were typically operated at or very close to the convenient temperature of liquid helium, $T\approx 4.2$~K. The first reported temperature-dependent measurements \cite{Engel04a,Semenov05,Kitaygorsky05,Kitaygorsky07} were focussing on dark-count rates, which can be significantly reduced by lowering the temperature, just as it is expected for thermally-activated fluctuations. Soon afterwards, the first reports were published about the temperature-dependence of the photon detection in the visible and near-infrared spectral range of NbN \cite{Rosfjord06,Goltsman07a,Kerman07}, NbTiN \cite{Dorenbos08}, and Nb SNSPD \cite{Annunziata09}. All of these studies have confirmed the following general trends: reducing the operation temperature of the detector from $4$ to $2$~K shifts the cut-off wavelength, above which the detection efficiency drops fast with increasing wavelength, to longer wavelengths, and the detection efficiency at wavelengths shorter than the cut-off wavelength increases to a certain extent or stays at least constant. These two effects combined lead to a significantly increased detection efficiency at the wavelengths of $1.3~\mu$m and $1.55~\mu$m, which are the two most important wavelengths for optical communication applications.

In the simple hot-spot model \cite{Semenov01}, as well as in a model taking into account the reduction of the critical current by excess quasi-particles \cite{Semenov05a}, the cut-off wavelength strongly depends on the superconducting energy gap: $\lco\propto\Delta^{-2}$. Recently, it was experimentally verified that a reduced energy-gap $\Delta$ leads to an increased cut-off wavelength in SNSPD made from WSi \cite{Baek11} and TaN \cite{Engel12}, two materials with a smaller intrinsic energy-gap than in NbN. Analogously, one would expect an increase of the cut-off wavelength with a decreasing energy gap for increasing temperature, which is in contrast to experimental observations. Despite this apparent discrepancy between experiments and theoretical models, no study has been published so far attempting to explain or study in detail this puzzling discrepancy. In the following, we report on systematic measurements of the temperature-dependence of the spectral detection efficiency of TaN and NbN SNSPD. In addition to the increase of the cut-off wavelength and the increase of the detection efficiency at lower temperatures, we have observed a change in the wavelength-dependence of the detection efficiency for wavelengths longer than the cut-off wavelength. Theoretical models will be discussed that might lead to a better understanding of the experimental observations.

\begin{table*}[!t]
\renewcommand{\arraystretch}{1.3}
\caption{Relevant parameters for the SNSPD investigated in this study. Definitions and details are given in the text.}
\label{tab.parameters}
\centering
\begin{tabular}{lccccccccc}
\hline
 & $d$ (nm) & $w$ (nm) & $T_c$ (K) & $\rho_N\ (\mu\Omega$cm) & $\xi_0$ (nm) & $\lambda_0$ (nm) & $D_e$ (cm$^2$/s) & $I_{b_1}(T)/I_c(T)$ & $I_{b_2}(T)/I_c(T)$\\
\hline
TaN1 & $4.9$  & $110$  & $9.3$  & $2.1$  & $5.0$  & $490$  & $0.60$ & $0.36$  & $0.41$\\
TaN2 & $4.9$  & $110$  & $8.8$  & $2.2$  & $5.2$  & $520$  & $0.58$ & $0.45$  & $0.50$\\
NbN1 & $3.8$  & $100$  & $10.9$  & $2.0$  & $4.5$  & $440$  & $0.50$ & $0.46$  & -- \\
\hline
\end{tabular}
\end{table*}

\section{Experimental details}

Thin films of TaN and NbN with thicknesses $d\approx4-5$~nm were prepared by DC magnetron sputtering in a N$_2$/Ar-atmosphere on R-plane cut sapphire substrates \cite{Ilin08,Ilin12a}. The substrates were kept at $\approx750^\circ$C during deposition. Partial gas pressures and deposition rates were chosen to obtain films with the maximum material-dependent critical temperature $T_c$. The films were structured into the standard meander geometry using e-beam and photo-lithography (for larger structures) and subsequent reactive ion-etching. We estimated the film thicknesses $d$ from pre-determined deposition rates and deposition times, and the conduction path widths $w$ were determined from scanning electron microscope images.

The relevant superconducting parameters for photon detection were deduced from conductivity measurements as a function of temperature. A detailed analysis \cite{Bartolf10} of conductivity data allowed us to obtain reliable values for $T_c$, normal-state resistivity $\rho_N$ at $T_c$, zero-temperature coherence length $\xi_0$, penetration depth $\lambda_0$, and normal electron diffusion coefficient $D_e$ at $T_c$. These values are given in Table \ref{tab.parameters} for the devices studied.

Photon-count rates as a function of photon wavelength $\lambda$ were measured in either a $^4$He-bath cryostat or a $^3$He-bath cryostat. The temperature-range in the $^4$He setup was limited from $\sim 4.5$~K to $6$~K, whereas in the $^3$He system a much broader temperature-range from $\approx0.5$~K up to $10$~K was accessible. Both free-space setups used an incandescent light source emitting a continuous light spectrum. The light was passed through a monochromator and a polarizer before being directed through a series of quartz windows and onto the detector. The detectors were biased with stable DC current-sources, and the detector voltage-pulses were amplified by broadband microwave amplifiers for detection by a threshold-level counter. At all temperatures the current dependencies of the count rate (examples are shown in the inset of Fig.\ \ref{Fig.T-model}) clearly show the plateau indicating the single-photon detection regime.

\section{Results and Discussion}

Currently, there are two slightly different models that are commonly used to describe the photon-detection in SNSPD \cite{Semenov01,Semenov05a}. In both models, the minimum photon energy that can be detected depends on the applied bias current relative to the depairing critical-current $I_c(T)$ of the superconducting strip. It is therefore important to measure at a constant reduced bias-current $I_b/I_c(T)$. However, it is inherently difficult to measure $I_c(T)$ with the required accuracy over the full temperature-range of interest \cite{Ilin12a}. Instead, we use here the approach by Bardeen\cite{Bardeen62} to calculate the temperature-dependent critical current
\begin{equation}\label{Eq.CritCurr}
I_c(T) = \frac{4\pi^{5/2}}{21\zeta(3)\sqrt{3}}\frac{(k_BT_c)^{3/2}}{e\rho_N\sqrt{D\hbar}}wd\left[1-\left(\frac{T}{T_c}\right)^2\right]^{3/2},
\end{equation}
with $\zeta(3)=1.202$, $k_B$ the Boltzmann-constant, $e$ the elementary charge, and $wd$ the cross-sectional area of the conduction path. It is worth noting that this approach is applicable only if the relevant current scale is indeed the depairing critical-current or another current-scale with a similar temperature dependence, nevertheless we will identify the critical current with the depairing critical-current given by Eq.\ (\ref{Eq.CritCurr}). Furthermore, the critical currents calculated with Eq.\ (\ref{Eq.CritCurr}) are typically about a factor of 2 larger than experimentally measured ones. Reduced bias currents $I_b/I_c$ are correspondingly smaller compared to most other reported data, where bias currents are scaled using the experimental critical current.

\begin{figure}[!t]
\centering
\includegraphics[width=\columnwidth]{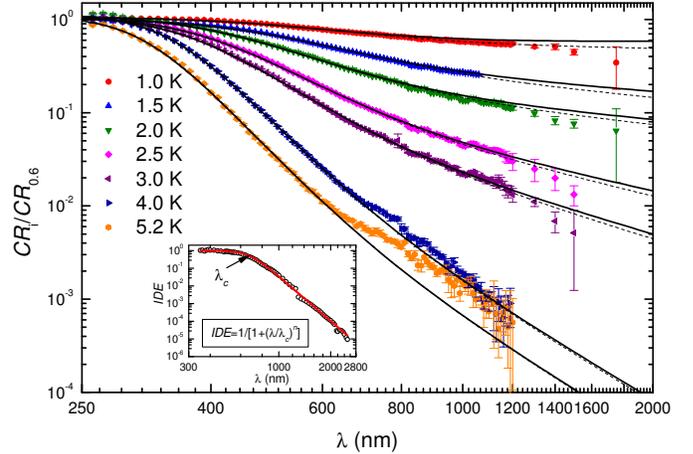}
\caption{Relative photon-count rates at different temperatures as indicated in the graph as functions of photon wavelength and constant $I_b(T)/I_c(T)=0.36$ (TaN1) in a double-logarithmic plot. The reference measurement was taken at $T=0.6$~K. Thin dashed-lines are least-squares fits to the data, solid lines are calculated using Eq.\ (\ref{Eq.relCR}) and best estimates for the parameters as described in the text. The inset shows a measurement using a full calibration of the optical setup, and a fit according to Eq.\ (\ref{Eq.DElambda}) (red line).}
\label{Fig.relCR}
\end{figure}

We used an empirical power-law to describe the spectral dependence of the detection efficiency
\begin{equation} \label{Eq.DElambda}
DE(\lambda) = DE_T\frac{1}{1+(\lambda/\lco)^n}.
\end{equation}
Here, $DE_T$ is the potentially temperature-dependent detection efficiency in the limit of short wavelengths (the so-called \emph{plateau region}) and the exponent $n$ is an empirical parameter describing the fast drop of the detection efficiency for long wavelengths. With Eq.\ (\ref{Eq.DElambda}) or variants thereof\cite{Lusche13} we can successfully describe the wavelength-dependence of $DE$ over several orders of magnitude in $DE$. In the inset of Fig.\ \ref{Fig.relCR} we show the intrinsic detection efficiency $IDE$ for the detector TaN1 at $5.2$~K and at a particular bias current. In order to become insensitive to the calibration of the photon flux-density and the distinction between intrinsic and device detection efficiencies, we went one step further and considered the wavelength-dependence of relative photon-count rates $CR_{T_2}(\lambda)/CR_{T_1}(\lambda)$ measured for two different temperatures $T_1$ and $T_2$. In fact, in this way we do not need any information about a possible spectral dependence of the optical coupling efficiency or any absolute calibration of the photon flux density, but only have to monitor and correct for variations of the lamp intensity. Such relative photon-count rates can then be described by
\begin{equation} \label{Eq.relCR}
\frac{CR_{T_2}(\lambda)}{CR_{T_1}(\lambda)}=\frac{DE_{T_2}}{DE_{T_1}}\frac{1+(\lambda/\lambda_{c,1})^{n_1}}{1+(\lambda/\lambda_{c,2})^{n_2}},
\end{equation}
where we choose $T_2>T_1$, such that typically $CR_{T_2}(\lambda)/CR_{T_1}(\lambda)\leq1$. With this approach, we are able to determine the temperature dependence of the cut-off wavelength \lco, the exponent $n$, as well as the relative detection efficiency.

\begin{figure}[!t]
\centering
\includegraphics[width=\columnwidth]{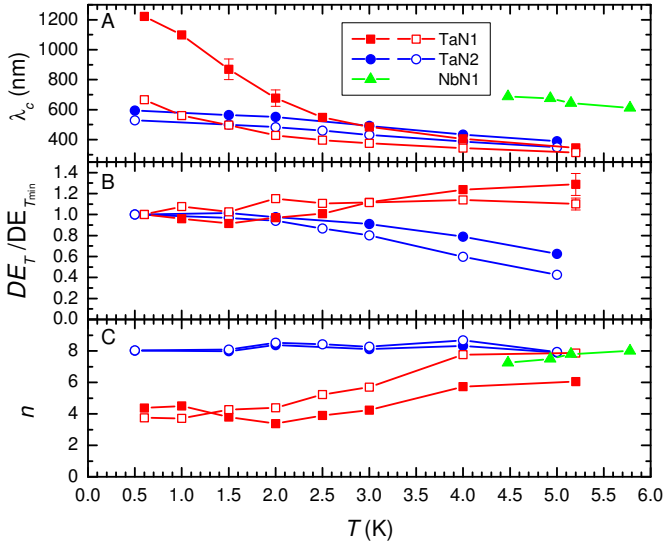}
\caption{Temperature dependence of cut-off wavelength (A), relative detection efficiency (B), and power-law exponent (C) for all detectors. The TaN SNSPD have been measured for two different relative bias-currents $I_b(T)/I_c(T)$ (see Tab.\ \ref{tab.parameters}, open symbols are for the lower current $I_{b_1}$, and filled symbols are for the higher current $I_{b_2}$). The relative detection efficiencies of the NbN SNSPD were fixed to $1$ for fitting the data and are not plotted.}
\label{Fig.T-dep}
\end{figure}

In Fig.\ \ref{Fig.relCR} we show a set of relative photon-count rates for various temperatures with $T_1=0.6$~K being the reference measurement. Least-squares fitting of Eq.\ (\ref{Eq.relCR}) to the data (dashed lines in Fig.\ \ref{Fig.T-dep}) allowed us to retrieve a set of fitting parameters, in which the parameters at the reference temperature are the same for all curves and are temperature-dependent fitting parameters at the higher temperatures. We repeated this procedure, sequentially taking the data at the next higher temperature as a new reference measurement and dropping the lowest temperature. In this way we obtained a certain number of best-fit values for each fitting parameter at the measurement temperatures. We then took the average of these values for each temperature as best estimates for the parameters entering Eq.\ (\ref{Eq.relCR}) (solid lines in Fig.\ \ref{Fig.relCR}). The resulting temperature-dependence of $\lco$, $n$, and the relative detection efficiency are plotted in Fig.\ \ref{Fig.T-dep} for the reduced currents given in Table\ \ref{tab.parameters}.

\subsection{Cut-off Wavelength} \label{Sec.cut-off}

The cut-off wavelengths for all 3 detectors (panel A in Fig.\ \ref{Fig.T-dep}) show the typical trend of increasing \lco\ with decreasing temperature. However, there is a marked difference between TaN1 and TaN2, which nominally should be very similar. The temperature-dependencies of the other parameters (panels B and C in Fig.\ \ref{Fig.T-dep}) are also different. Most likely these observations can be explained by the different age of the detectors. Whereas TaN1 has been measured shortly after production, TaN2 was measured many months after it was produced. Although it had been stored under a rough vacuum, it seems that thin TaN films are susceptible to long-term structural and/or chemical changes on a time-scale of several months.

\begin{figure}[!t]
\centering
\includegraphics[width=\columnwidth]{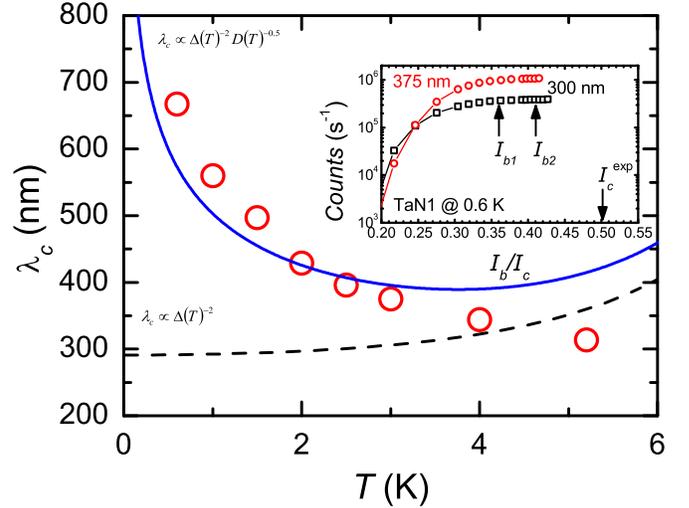}
\caption{Temperature dependence of the cut-off wavelength for TaN1 at $I_b(T)/I_c(T)=0.36$, in comparison with expectations from the simple hot-spot model (Eq.\ (\ref{Eq.simple}), dashed, black line) and the quasi-particle diffusion model (Eq.\ (\ref{Eq.diffusion}), solid blue line). Both model predictions have been shifted vertically and are not the result of a fitting procedure. The inset shows photon count rates vs.\ bias current $I_b/I_c$, applied bias currents for spectral measurements and the experimental critical current are indicated.}
\label{Fig.T-model}
\end{figure}

In the simple hot-spot model that assumes a normal-conducting core, which diverts the bias-current into smaller cross-sections on either side of the core, the cut-off wavelength should be \cite{Maingault10}
\begin{equation} \label{Eq.simple}
\lco(T) \propto \left(N_0\Delta^2\right)^{-1}\left(1-\frac{I_b(T)}{I_c(T)}\right)^{-2},
\end{equation}
with $N_0$ the density of states at the Fermi-level (assumed to be temperature-independent), and all other factors being constants or given by the geometry. Since we have kept $I_b(T)/I_c(T)$ constant, the only temperature-dependent quantity remaining is the superconducting gap $\Delta(T)=\Delta_0(1-t^2)^{0.5}(1+t^2)^{0.3}$ with $t=T/T_c$, approximating $\Delta(T)$ for all $T\leq T_c$. The resulting expectation for \lco\ is plotted in Fig.\ \ref{Fig.T-model} (dashed line), which is clearly incompatible with the experimental data for TaN1 at $I_b(T)/I_c(T)=0.36$, and similar inconsistencies are observed for the other data-sets as well.

Taking into account diffusion of quasi-particles that are created after photon absorption, and the fact that excess quasi-particles lead to a reduction of the critical-current, one arrives at a slightly different relation for the cut-off wavelength (for more details see Ref.\ \cite{Semenov05a})
\begin{equation}\label{Eq.diffusion}
\lco \propto \left[N_0\Delta^2\sqrt{D_q\tau}\left(1-\frac{I_b(T)}{I_c(T)}\right)\right]^{-1},
\end{equation}
where the thermalization time $\tau$ and the diffusion coefficient $D_q$ of the quasi-particles may, in principle, be temperature dependent. In a first approach we kept $\tau$ constant and only allowed for a temperature dependence of $D_q$. We have numerically calculated $D_q(T)$, for $T<T_c$, using the general relation $D=\kappa_e/c_e$ with BCS expressions for the electronic thermal conductivity $\kappa_e$ \cite{Abrikosov88}, and the electronic specific heat $c_e$ \cite{Tinkham96}. The resulting temperature-dependence $\lco(T)$ is plotted in Fig.\ \ref{Fig.T-model} (solid blue line). It is obvious that this approach reproduces the significant increase in \lco\ for $T<2$~K very well. We note that the calculated $D(T)$ is only valid for thermalized quasi-particles. At temperatures closer to $T_c$, the strong temperature-dependence of the superconducting gap should dominate and lead again to an increase in $\lco$. Unfortunately, we were not able to extend our measurements to higher temperatures, due to rapidly decreasing critical-currents, resulting in small bias currents and therefore small signal amplitudes. Also, the detectors became increasingly unstable at elevated temperatures.

It is interesting to note that the recently developed X-ray superconducting nanowire single-photon detectors (X-SNSPD) \cite{Inderbitzin12} seem to show a different temperature dependence of the minimum detectable photon energy. Results obtained for a $100$~nm-thick TaN X-SNSPD indicate a decreasing threshold energy (\emph{i.e.} increasing \lco) with increasing operation temperature \cite{Inderbitzin13}.

\subsection{Relative Detection Efficiency}

In panel B of Fig.\ \ref{Fig.T-dep} we show the relative detection efficiency versus temperature. It is remarkable that the detection efficiency in the plateau region remains essentially constant for the high-quality detector TaN1, whereas the somewhat degraded detector TaN2 shows a pronounced reduction of the detection efficiency with increasing temperature. The exact reason for this behavior is unclear at this stage, but we expect that an increased level of inhomogeneities plays an important role here. Measuring the detection efficiency as a function of temperature may therefore be a tool to determine a the quality of a detector.

\subsection{Power-Law Exponent}

In panel C of Fig.\ \ref{Fig.T-dep} we plot the power-law exponents $n$ in Eq.\ (\ref{Eq.relCR}) as a function of temperature. This quantity describes the decrease of the detection efficiency beyond the cut-off wavelength \lco\ for photons with energies that are insufficient to trigger a normal-conducting domain in the superconducting strip without any additional contribution from fluctuations. A thermally-activated crossing of vortices has been suggested as a possible mechanism to describe the detection of such photons \cite{Bulaevskii12}. However, recent measurements of the magnetic-field dependence of photon-count rates raised serious doubts about the role of vortices in this regime\cite{Engel12a}.

Interestingly, the power-law exponents of the degraded TaN2 detector are consistent with a temperature independent behavior, while the data of the high-quality TaN1 and NbN1 SNSPD suggest an increasing exponent with increasing temperature. At this point of our investigations we have no explanation for such a behavior, but we believe that it may give helpful clues to uncover the true mechanism of low-energy photon detection in these devices.

\section{Conclusion}
We have presented systematic measurements of the temperature-dependence of the spectral detection-efficiency of SNSPD. We have tried to eliminate the temperature-dependence of the critical current by measuring at a constant reduced bias-current $I_b(T)/I_c(T)$, where we have used Bardeen's result for the depairing critical current. We have found that the wavelength-dependence of the detection efficiency can be accurately described by a simple power-law, which in turn allowed us to extract not only the temperature dependence of the cut-off wavelength, but also the temperature dependencies of the detection efficiency for wavelengths shorter than \lco, as well as of the power-law exponent in Eq.\ (\ref{Eq.relCR}) for wavelengths $\lambda>\lco$.

We have confirmed that the cut-off wavelength increases with decreasing temperature and suggested a temperature-dependent diffusion coefficient as the main underlying cause. This observation is strong evidence for the important role of quasi-particle or energy diffusion for a correct understanding of the detection mechanism. These results may further stimulate theoretical developments of a microscopic detection model \cite{Zotova12}.

High-quality SNSPD, defined as detectors with a low level of inhomogeneities, are most likely characterized by a temperature-independent detection efficiency for $\lambda<\lco$. In that case the device detection efficiency appears to be given by the absorption probability and an intrinsic detection efficiency close to unity. Detectors with non-negligible inhomogeneities can still reach high detection efficiencies by operating them at low temperatures.

The fluctuation-enhanced detection of photons with $\lambda>\lco$ remains an interesting topic. We have found that in this regime the decrease in detection efficiency can be well described by a power-law, $DE\propto\lambda^{-n}$, over many orders of magnitude in $DE$. Furthermore, it seems that the reduction of the detection efficiency with increasing photon wavelength is slower at low temperatures, particularly in high-quality devices. Any correct description of the detection mechanism in this regime will have to be able to explain these observations.


\end{document}